\documentclass[pra,twocolumn]{revtex4-1}
\usepackage{graphicx, amsmath, amsfonts, amssymb}
\begin{document}
\title{Multi-slit interferometry and commuting functions of position and momentum}
\author{Johannes CG Biniok}
\email{jcgb500@york.ac.uk}
\affiliation{University of York, York YO10 5DD, UK} 
\author{Paul Busch}
\email{paul.busch@york.ac.uk}
\affiliation{University of York, York YO10 5DD, UK} 
\date{\today}

\begin{abstract}
In a recent, modified double-pinhole diffraction experiment the existence of an interference pattern was established indirectly along with a near-perfect imaging of the double pinhole. Our theoretical analysis shows that the experiment constitutes a preparation of a quantum state that is, to a good approximation, a joint eigenstate of commuting functions of position and momentum. Gaining information about the momentum distribution by means of the particular experimental setup is thus possible with negligible impact on the position distribution. Furthermore, we construct explicitly a class of states simultaneously localised on periodic sets in position and momentum space, which are therefore eigenstates of the observables being measured jointly (to a good approximation) in multi-slit interferometry. Finally, we show that with an appropriate change of settings the experiment demonstrates the mutual disturbance of position and momentum measurements.
\end{abstract}
\maketitle

\section{Introduction}
Still at the heart of quantum mechanics, the double-slit experiment remains the subject of ongoing 
investigation with surprising results that attract wide attention. For example, a recent interference experiment 
\cite{Steinberg2011} exhibited ``average trajectories'' via weak measurements. Here, we revisit
another experiment, reported in 2007 \cite{Afshar2007}, which investigates the influence of a wire grating placed
at the nodes of the interference pattern on the image of the double-pinhole. While the authors' own theoretical account seems untenable and has been criticised, the experiment itself remains interesting. As we argue here, in this experiment a quantum state is prepared that is approximately a joint eigenstate of position and momentum on periodic sets, and verified with negligible disturbance. 

Multi-slit interference experiments, such as Young's double-slit experiment, consist of a coherent source, an aperture mask and a detection screen (placed in the far-field). Within the framework of quantum mechanics, such an experiment is viewed as follows: While the aperture mask prepares a particle in a quantum state with a certain position distribution, the observed interference pattern is a measurement of the associated momentum distribution.

Traditionally, the momentum distribution is captured on a detection screen, but this clearly destroys the quantum state. Establishing the existence of an interference pattern indirectly, i.e. without destroying the quantum state, is possible by removing the screen and replacing it by a wire grating, each wire carefully placed at the location of a node in the interference pattern \cite{Afshar2007}. The existence of an interference pattern may be deduced from the practically undiminished intensity passing the wire grating. Using a lens, a geometric image of the aperture is produced, which allows detection of the quantum particle on the very set of positions it was prepared on -- after it was subjected to a momentum measurement. While indirectly observing an interference pattern without changing the localisation properties of a system may not be surprising from the point of view of classical physics, it is rather curious when considered in terms of quantum mechanics: Information about a quantum state was obtained, but apparently without changing the properties of that quantum state. In particular, information about a pair of incompatible observables was obtained; in this context, the measurement seems `classical', revealing already existing information without changing the system properties.

This observation indicates that the experiment should be described in terms of two commuting observables which yield information about position and momentum respectively. While position and momentum do not commute, functions of position may commute with functions of momentum. Indeed, as will be shown here, the experiment can be considered an approximate realisation of a joint eigenstate of mutually commuting functions of position and momentum. In the next two sections, the experimental setup and joint eigenstates of periodic sets of position and momentum are discussed. This is followed by a description of multi-slit experiments in terms of joint eigenstates. 

The experiment reported in \cite{Afshar2007} was performed with photons; its analysis would require a treatment in terms of photons as massless spin-1 particles, which are known to be only unsharply localisable. (A review of the problem of photon localisation and relevant literature where unsharp localisation observables for the photon are introduced can be found in \cite{OQP}.) For simplicity, the treatment here is non-relativistic and strictly only applies to matter waves. There is nevertheless good qualitative and quantitative agreement between our theoretical analysis and the experiment, suggesting that an analogue of the non-relativistic argument applicable to photons should exist, and showing that the experiment demonstrates a simultaneous determination of compatible, coarse-grained versions of the complementary position and momentum observables.

\section{On the experimental setup} 
\begin{figure}
	\centering 
	\includegraphics[width=0.5\textwidth]{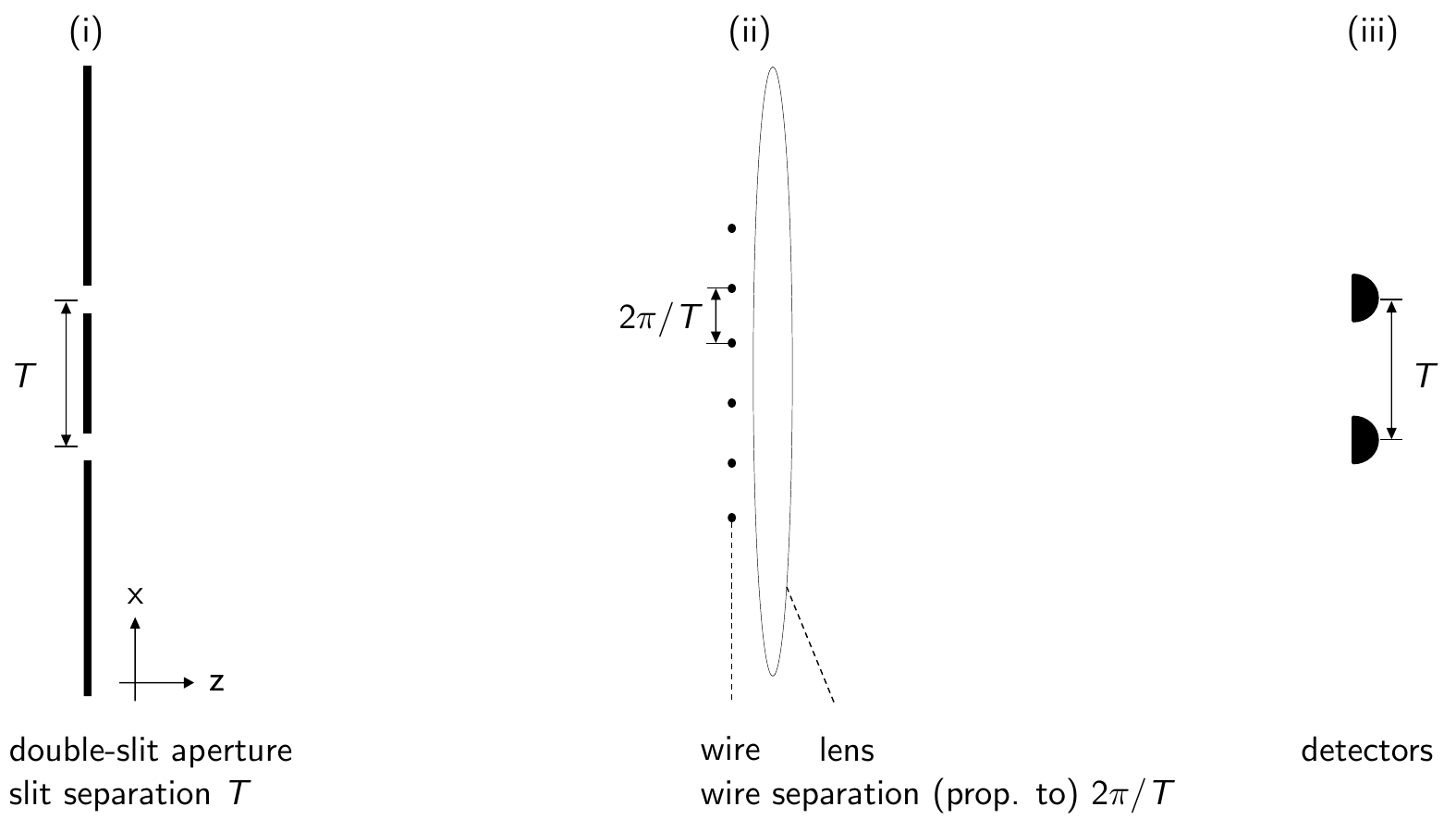} 
	\caption[]{Simplified illustration of the setup used to indirectly establish the existence of an interference 
	pattern of a coherently illuminated double-slit aperture.} 
	\label{fig:Diagram} 
\end{figure}

The setup illustrated in Figure \ref{fig:Diagram} depicts a simplified version of the experiment reported in \cite{Afshar2007}. While the experiment was performed using a double-pinhole, here a double-slit setup is considered. A particle propagates through the device along the $z$-axis (from left to right). We model its wavefunction as a product, $\Psi(x,y,z)=\phi(x)\eta(y)\zeta(z)$, and focus on the component $\phi(x)$, where the $x$-axis is along the transversal (vertical) direction. 
The state $\zeta(z)$ is a means of keeping track of the times of passage through the experimental setup. As is detailed in Appendix A and used later on, in the appropriate limit this problem can be simplified so that only $\phi(x)$ needs to be considered, removing any explicit time dependence but retaining an identification of the quantum state at different times with distinct locations in the setup. 

The wavefunction $\phi(x)$ is diffracted at location (i), where the double-slit aperture mask is depicted. A wire grating is placed at location (ii), where the interference pattern would be expected. The separation of the wires depends on the spacing of the slits in the aperture mask via the indicated reciprocal correspondence $T\leftrightarrow 2\pi/T$, although in general the wavelength of the source must be taken into account.
A lens is placed immediately behind the wire grating for the purpose of producing the geometrical image of the original aperture at location (iii), where the detectors are placed. As is argued below, the sequential character of the setup, with the aperture at (i) and the grating at (ii), actually constitutes a joint preparation/measurement.

\noindent
The action of the aperture mask at (i) is modelled by the following transmission function that gives the wavefunction $\psi$ (up to normalisation) after passage through the aperture:
\begin{equation} \label{transF}
\phi(x) \to \chi_{A}(x)\,\phi(x) 
\equiv C \psi(x).
\end{equation}
Here $C$ is a normalisation constant and $\chi_{A}(x)$ is the indicator function of set $A$, with value $1$ for $x \in A$ and $0$ otherwise; $A$ being the
set that describes the effective aperture. Incidentally, Eq.~\eqref{transF} defines the action of an operator that is
defined as a function $\chi_A(Q)$ of the position operator $Q$:
\[ \left(\chi_A(Q)\,\phi\right)(x):=\chi_A(x)\phi(x).\]
This operator has eigenvalues 1 and 0 with associated eigenfunctions given by functions $\phi(x)$ either localised
within $A$ or within the complement of $A$. Thus,
 the state vector $\phi$ is projected onto an eigenvector
of the {\em spectral projector} $\chi_A(Q)$ of ${Q}$ associated with the set $A$.
For coherent illumination of both
slits, a wavefunction with two isolated peaks is
prepared. Such a superposition state is henceforth denoted $\psi_{2}$. A
single-slit wavefunction, denoted $\psi_{1}$, is used to describe a single-slit state.

The aperture mask at location (i) prepares the quantum state represented by the wavefunction $\psi$, which then propagates freely until it arrives at (ii). In the Fraunhofer limit, upon arriving at (ii) the wavefunction has evolved so as to have a profile
approximately identical (up to scaling) to that of the Fourier-transform $\widetilde{\psi}$ of the wavefunction at (i). 
{For more details, the reader is referred to Appendix A.}

The effect of the wire grating is modelled by a transmission function similar to the one specified in (\ref{transF}), but with a set $B$ of intervals complementing the regions occupied by the wire grating:
\begin{equation*} 
\widetilde\psi(k)=(\mathcal{F}\psi)(k) \to \chi_{B}(k)\,\widetilde\psi(k)\equiv \left(\chi_B( P)\,\widetilde\psi\right)(k),
\end{equation*}
where the arrow indicates passage through the wire grating and $\chi_B( P)$ denotes the spectral projector of momentum ${P}$ associated with the set $B$ and $\mathcal{F}$
denotes the unitary operator effecting the Fourier transform,
\begin{equation*} 
\widetilde{f}(k)=(\mathcal{F}f)(k)=\frac{1}{\sqrt{2\pi}}\int_{-\infty}^{\infty}f(x)\,e^{i\,k\,x}\,dx.
\end{equation*}

In the experimental setup of \cite{Afshar2007}, a total of six wires is used, each with a diameter of 0.127\,mm and a separation of 1.3\,mm. 
It should be noted that for a single-slit interference pattern the wire grating would not be in the exact centre, 
but shifted sideways by a small amount. (In the experimental setup reported in \cite{Afshar2007}, the wire grating is shifted by 0.250\,mm while the single-slit interference pattern is of the order of tens of millimetres.)

Finally, the action of the lens located at (ii) is modelled as spatial inversion, expressed by mappings 
${Q} \mapsto -{Q}$ and ${P} \mapsto -{P}$, for the position and momentum respectively. 
This corresponds to the unitary parity transformation $\mathcal{P}$, which coincides with the square of the 
Fourier transformation $\mathcal{F}$.
As a result, the divergent wave rays emerging, say, from the double pinhole and arriving at the wire grating and 
lens are inverted so as to be refocused into an image of the original double slit.

\section{Commuting functions of position and momentum}
While the canonical commutation relation $[{Q},{P}]=i $ (we will put $\hbar=1$ throughout) represents the fact that the position and momentum observables are incompatible in a strong sense, a function of position may commute with a function of momentum. 
A first characterisation of commuting functions of position and momentum was given in \cite{APP} in the context of an analysis of interference experiments, with the aim of explaining non-local momentum transfers in the Aharonov-Bohm effect. 
A first full proof of necessary and sufficient conditions for the commutativity of functions of position and momentum was
was reported in \cite{Busch1986}, who were unaware of the work of \cite{APP}. 
A first construction of a set of joint eigenstates was given in \cite{Reiter1989}. Here, we present a construction of joint eigenstates that is readily identified with multi-slit interferometry. In Appendix C an alternative, rigorous construction is included that generalises \cite{Reiter1989}.

Considering the commutation relation in a form due to Weyl, 
\begin{equation*} 
e^{i \, p \, {Q}}e^{i \, q \, {P}}=e^{-i \, p \, q}e^{i \, q \, {P}}e^{i \, p \, {Q}},
\end{equation*}
the existence of commuting functions of ${Q}$ and ${P}$ is suggested since the operators $e^{i \, p \, {Q}}$ and $e^{i \, q \, {P}}$ commute for $pq=2\pi n$ with $n\in \mathbb{N}$. Though ${Q}$ and ${P}$ do not commute, the spectral projections $\chi_X( Q)$ and $\chi_Y( P)$ for periodic sets $X$ and $Y$ commute if the sets have periods $T$ and ${2\pi}/(nT)$, respectively, where $n\in \mathbb{N}$:
\begin{equation*} 
\left[\chi_X( Q),\chi_Y( P)\right]=0.
\end{equation*}
(A set $X$ is called periodic with (positive minimal) period $T$, if $T$ is the smallest positive number by which $X$ can be shifted such that the shifted set $X+T=X$, or equivalently, if its indicator function is a periodic function with minimal period $T$.)

Physical systems exhibiting such doubly periodic behaviour occur naturally. A well known example is found in solid state physics: The wavefunction of an electron in a crystal is not only periodically localised in accordance with the periodic potential that is due to a crystal lattice; the wavefunction is also periodically localised in momentum space (this is encapsulated in the notion of the reciprocal lattice). While solid state physics often deals with systems containing a very large (essentially infinite) number of lattice points, even finite multi-slit experiments can be regarded as an approximate realisation of joint eigenstates of $\chi_X( Q)$ and $\chi_Y( P)$ over periodic sets as is argued below. 

The following construction of a class of joint eigenvectors is carried out using the Dirac comb $\Delta_T$, 
defined as 
\[\Delta_T(x)=\sum_{n=-\infty}^{\infty}{\delta(x-nT)},\]
where $\delta$ denotes the delta-distribution.
This has heuristic value and also makes the identification with multi-slit experiments more intuitive. 
We note that under a Fourier transformation the Dirac comb $\Delta_T$ with period $T$ becomes a Dirac comb 
with period ${2\pi/T}$:
\begin{equation} \label{DC}
\mathcal{F} \left(\Delta_{T}\right)(k) = \frac{1}{T} \;\Delta_\frac{2\pi}{T}(k).
\end{equation}

The sought joint eigenstates of $\chi_X( Q)$ and $\chi_Y( P)$ must have position and momentum representations
that are localised in the periodic sets $X$ and $Y$, respectively. Their construction makes use of 
the following identity involving functions $W$ and $M$ which will be suitably chosen:
\begin{equation} \label{diff3}
\mathcal{F} \Big(W \ast (\Delta_T\cdot M) \Big)(k) = \left(\widetilde{W}\cdot\left(\frac 1T\Delta_\frac{2\pi}{T} 
\ast \widetilde{M}\right)\right)(k).
\end{equation}
The order of the two operations in (\ref{diff3}), convolution ($\ast$) and multiplication, may be 
chosen freely, though the result is different in general. Here, both orders appear naturally because 
of the Fourier transformation present. (A special case of \eqref{diff3} is applied in \cite{Corcoran2004} for the 
construction of functions invariant under Fourier transformation.)

We now choose $W$ and $\widetilde{M}$ to be square-integrable functions that are localised on (that is, vanish exactly outside) intervals of lengths strictly less than $T$, resp. $2\pi/T$. (It will be convenient to use the mathematical term {\em support (of a function)} when speaking of the smallest closed set on which the function is localised.) This ensures that the function $\psi$ defined via (\ref{diff3}) is indeed square-integrable {(See Appendix B)}: 
\begin{align} 
\psi(x) &= \Big(W \ast (\Delta_T\cdot M) \Big)(x)\label{psi},\\
\widetilde{\psi}(k) &= \left(\widetilde{W}\cdot\left(\frac 1T\Delta_\frac{2\pi}{T} 
\ast \widetilde{M}\right)\right)(k).\label{psitilde}
\end{align}
The wavefunction $\psi$ is now localised on a periodic set $X$ with period $T$, and its Fourier transform $\widetilde{\psi}$ is localised on a periodic set $Y$ with period $2\pi/T$. These sets are indeed obtained by placing equidistant copies of the supports of $W$ and $\widetilde{M}$, respectively. It follows in line with the result of \cite{Busch1986} that $\psi$ is a joint eigenstate of the associated spectral projections of position and momentum.
A mathematically rigorous construction of such joint eigenstates without the use of Dirac combs is included in the Appendix B.

The vector $\psi$ thus does not change under the action of these spectral projections $\chi_X( Q)$ and $\chi_Y( P)$. In general, for any wavefunction $\phi$, the projected wavefunction\[\chi_Y( P)\chi_X( Q)\,\phi\]is a joint eigenstate of the two projectors. In fact, all eigenstates with eigenvalue 1 may be obtained as the projection onto the intersection of the ranges of $\chi_X( Q)$ and $\chi_Y( P)$, which is given by the product $\chi_X( Q)\chi_Y( P)=\chi_Y( P)\chi_X( Q)$. {In the analysis below we model the action of the slits and wires as projections in this sense.}

\section{Multi-slit experiments in terms of joint eigenstates of ${Q}$ and ${P}$ on periodic sets}\label{Sec:eigenfn}
As reported in \cite{Afshar2007}, an initial superposition state $\psi_2$ propagates through the experimental setup nearly undisturbed. By contrast, there is an effect on the image of the single-slit state $\psi_1$ detected at (iii): In addition to the expected intensity peak many smaller peaks are found, such that each peak is separated by a distance $T$ from its immediate neighbours. An illustration can be found in \cite{Afshar2007}, figures 1 c) and d) therein.

These two observations can be understood in terms of joint eigenstates of ${Q}$ and ${P}$ on periodic sets. Firstly, the superposition state $\psi_2$ remains unchanged to a good approximation, because $\psi_2$ is already prepared at (i) as a good approximation to a joint eigenstate of periodic characteristic functions of position and momentum with appropriate periodic sets $X,Y$. This makes $\psi_2$ an approximation to an eigenstate of the momentum projector associated with the opening left by the wire grating, and hence leaves it virtually undisturbed in the presence of the grating. This can be described symbolically by the approximate equations
\begin{equation*}
\psi_2 =\chi_X(Q)\psi_2 \to \chi_B(P)\psi_2 \approx \chi_Y(P)\psi_2 = \psi_2' \approx \psi_2.
\end{equation*}
Here $\psi\approx\phi$ is taken to mean $\|\psi-\phi\|\ll1$ for (sub-)normalised vectors, the arrow denotes passage through the wire grating.

Secondly, the single-slit state $\psi_1$ does not remain unchanged, but instead is detected on a set of locations expected of a joint eigenstate defined on a periodic set, as described above. It follows that the wire grating imposes nodes in a manner that approximates the action of $\chi_Y( P)$ to a high degree, because $\psi_1$ remains an eigenstate of $\chi_X( Q)$ after this action:
\begin{equation*} 
\psi_1=\chi_X( Q)\psi_1 \to 
\chi_Y( P)\psi_1=\psi_1' 
= \chi_X( Q)\psi_1'\not\approx \psi_1.
\end{equation*}
Considering that the experimental setup in \cite{Afshar2007} involves merely six wires, this may seem surprising. Without further analysis of the experimental details, this suggests that the part of the wavefunction not penetrating the wire grating must have comparatively small amplitude. This is elaborated below.

While all quantum states that pass the aperture mask are eigenstates of $\chi_X( Q)$, the combined effect of aperture and wire grating represents a preparation procedure for approximate joint eigenstates of $\chi_X( Q)$ and $\chi_Y( P)$: all quantum states are projected into the range of $\chi_X( Q)\chi_Y( P)=\chi_Y( P)\chi_X( Q)$ to a good approximation. The superposition state $\psi_2$ is thus already an approximate eigenstate of both projections, and the effect of the wire grating is much smaller than on the single-slit state $\psi_1$, and even negligible to a good accuracy.

Using (\ref{psi}) and (\ref{psitilde}), we now proceed to the construction of an example of a joint eigenstate of commuting periodic functions of ${Q}$ and ${P}$ that describes the double-slit setup. 

\begin{figure}
	\centering \vspace{6pt}
	\includegraphics[width=0.5\textwidth]{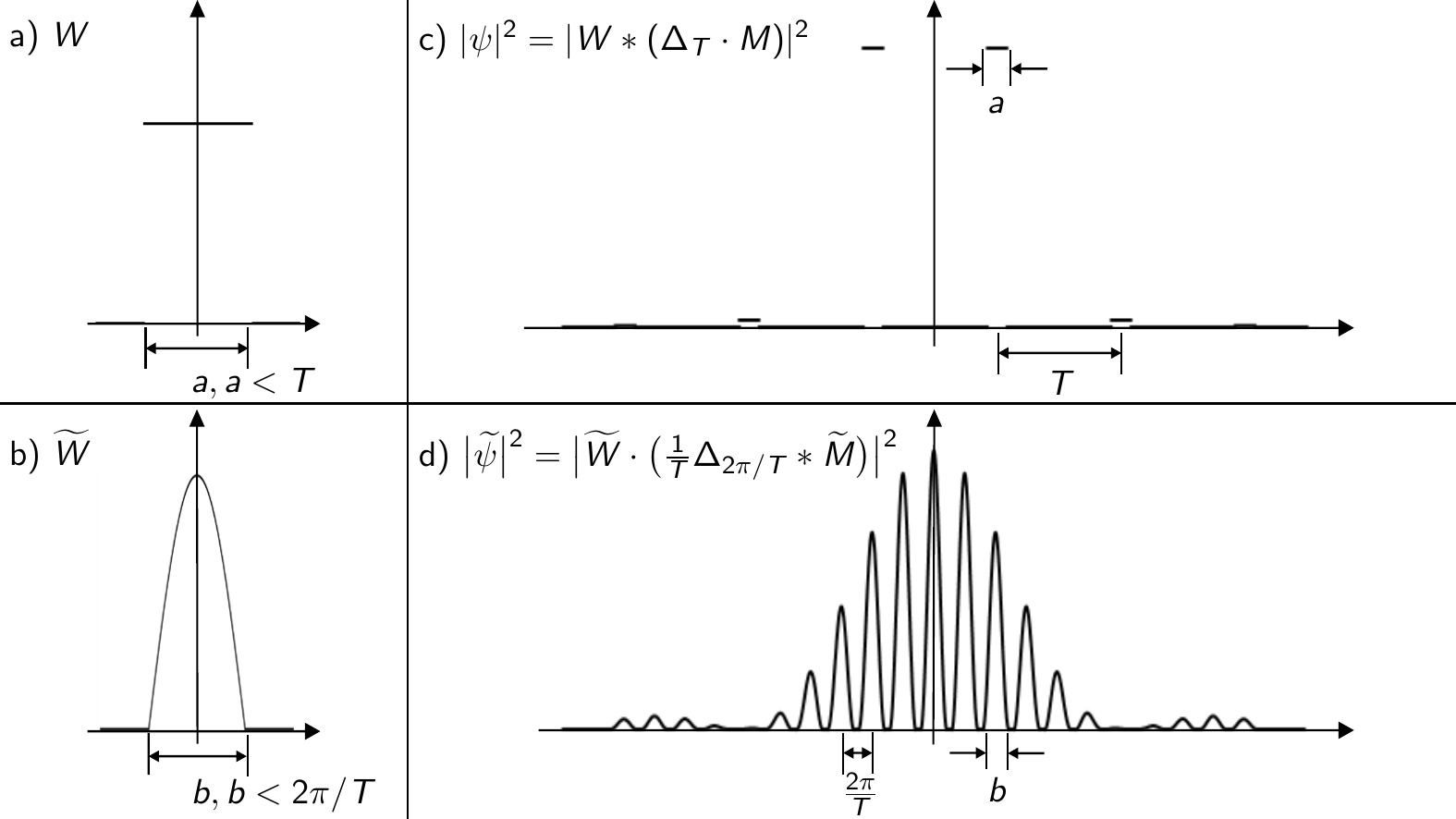} 
	\caption[]{Illustration of a state localised on periodic sets in position and momentum space. For this particular choice of $W$, $\widetilde{M}$, the tails of $|\psi|^2$ are negligible outside the two central slits, so that $X$ is well approximated by a double slit, and the negative-outcome testing of the localisation of $|\widetilde{\psi}|^2$ within $Y$ does not require many more than ten wires.} 
	\label{fig:Illustration} 
\end{figure}

\noindent
For this, the two localised functions $W, \widetilde{M}$ need to be chosen appropriately (Fig. 2). The function $W$ describes the quantum amplitude contained in a single slit. We consider the function that is constant on a single slit (a rectangular shape):
\begin{equation} \label{W}
 W(x)=\chi_{[-a/2,a/2]}(x)=
 \begin{cases}
  1 &\text{for $x$ $\in$ [$-a/2,a/2$]} \\
  0 &\text{for $x$ $\notin$ [$-a/2,a/2$]}
 \end{cases}
\end{equation}
Here, $a$ is the width of the slit. This is illustrated in Figure \ref{fig:Illustration} a). The expression for the Fourier
transform of $W$ is
\begin{equation*}
\widetilde{W}(k) \propto \mathrm{sinc}{(a k /2)}\,.
\end{equation*}
The function $\widetilde{W}$ accounts for the modulation of the interference
pattern $I(k)$. In the case of a double slit interference experiment with slit separation $T>a$, 
the known interference pattern $I_{ds}(k)$ is of the form
\begin{equation*}
I_{ds}(k) \propto \mathrm{sinc}^{2}(a k/2)\;\mathrm{cos}^{2}(T k/2).
\end{equation*}
The cosine describes a repeated pattern, and it suggests that we choose $\widetilde{M}$ to
correspond to a single instance of this pattern (in practice, this
choice would be made based on experimental data):
\begin{align} \label{M}
\widetilde{M}(k) & = \cos(T'k/2)\,\chi_{[-\pi/T',\pi/T']}(k)\nonumber \\
& = 
 \left\{ \begin{matrix}
  {\cos}(T'k/2) &\text{for $k\in [-\pi/T',\pi/T']$} \\
  0 &\text{for $k\notin[-\pi/T',\pi/T']$} 
 \end{matrix}\right..
\end{align}
This is illustrated in Figure \ref{fig:Illustration} b). For the $2\pi/T$-periodic set 
\[
Y=\bigcup_{n=-\infty}^\infty[2\pi n/T-\pi/T',2\pi n/T+\pi/T']
\] 
to be different from the 
whole real line, it is required that $T'>T$, so that the interval $[-\pi/T',\pi/T']$ is strictly 
contained in the interval $[-\pi/T,\pi/T]$. 

Combining the expressions obtained for $\widetilde{W},\widetilde{M}$ the interference pattern is described by:
\begin{widetext}
\begin{equation} \label{qEx2}
\left|\widetilde{\psi}_{2}(k)\right|^{2} \propto \frac{1}{T^2}\;\mathrm{sinc}^{2}(ak/2)\,\sum_{n=-\infty}^\infty\cos^2\left(\frac{T'}2\left(k+\frac{2\pi n}T\right)\right)\,
\chi_{[-\pi/T',\pi/T']}\left(k+\frac{2\pi n}T\right).
\end{equation}
\end{widetext}
This is a sum of non-overlapping terms, and the support of this function is the periodic set $Y$ that is made up of equidistant copies of the interval $[-\pi/T',\pi/T']$. For the quantum state in position space, $W$ is as defined in (\ref{W}), and $M$ follows from $\widetilde{M}$ as defined in (\ref{M}):

\begin{widetext}
\begin{equation} \label{ExaPsi}
\psi_{2}(x) \propto \Big(W \ast \big(\Delta_T((\,\cdot\,)-T/2) \cdot M\big) \Big)(x)=\ \sum_{n=-\infty}^\infty M((n-1/2)T)\,\chi_{[(n-1/2)T-a/2,(n-1/2)T+a/2]}(x).
\end{equation}\end{widetext}
The Dirac comb is shifted by $T/2$, in correspondence with the experimental setup. 
(This shift becomes a phase factor in momentum space and does not affect the momentum distribution.) 
Figures \ref{fig:Illustration} c) and d) illustrate $\psi$ and $\widetilde{\psi}$ as constructed 
in (\ref{ExaPsi}) and (\ref{qEx2}), respectively.

There are two important limiting cases. The spectral projection $\chi_Y( P)$ is over a strictly periodic set $Y$. In contrast, the dimensions of any experiment are necessarily finite, and, in particular, the experiment reported in \cite{Afshar2007} was performed with a total of six wires only, preparing the state $\chi_B(P)\psi$, where $B$ is the complement to the region occupied by the wires. A model calculation shows that the difference between the states $\chi_B(P)\psi$ and $\chi_Y(P)\psi$ is undetectable given the accuracy of the experiment at hand.

Finally we may consider the limiting case where $T'\to T$. This corresponds to the wires becoming negligibly thin. When $T'=T$ the function $M$ is zero at every delta peak of the periodic Dirac comb, except for two locations: $x=-T/2$ and $x=+T/2$. Hence it follows that for this particular choice of $W, \widetilde{M}$, the quantum state $\psi_2$ exists solely in the two slits and is an approximation to a joint eigenstate defined on periodic sets, where the wires must be assumed to have a very small thickness. The experimental setup considered prepares this quantum state at the aperture at location (i) as an eigenstate of $\chi_X( Q)$ for the periodic set $X$. Passage through a periodic wire set $Y$ will cause a projection of the state onto one that is a proper joint eigenstate of periodic position and momentum sets. This projective measurement action causes a disturbance of the incoming 2-slit wavefunction, which manifests itself in the observed position distribution at (iii): In an ideal setup with dimensions identical to those reported in \cite{Afshar2007}, 1\% of the total probability would not be found in the two detectors at (iii) where it would otherwise be expected. Instead, this one per cent of probability would be distributed over the remainder of the periodic set $X$. According to \cite{Afshar2007}, for the double-pinhole setup about $2\%$ probability were found outside of the main peaks.

\section{Discussion and Outlook}
A description of multi-slit experiments was presented, and in particular of the modified double-slit experiment in \cite{Afshar2007}, in terms of quantum states that are defined on periodic intervals of position and momentum. These quantum states, themselves not periodic, represent a class of joint eigenstates of periodic functions of position and momentum. Using a description in terms of such joint eigenstates it was possible to account for the two observations reported in \cite{Afshar2007} concerning the behaviour of a double-slit input state $\psi_2$ and a single-slit input state $\psi_1$. 

Firstly, an incoming double-slit superposition state is virtually unaffected by the indirect measurement of the interference pattern performed by the wire grating, with each of the wires placed at a node. This is, of course, because the superposition state shows an interference pattern. An explanation in terms of joint eigenstates over periodic sets, though, goes further and makes it possible to explain why a superposition state can be localised on essentially the same set of positions after it was subjected to such a measurement -- after all, measuring the interference pattern corresponds to measuring the momentum distribution. The information about position and momentum of the superposition state is approximately represented by commuting observables. It follows that there is no conflict with the principle of complementarity. The experimental setup constitutes a good approximation to a joint determination of compatible coarse-grainings of position and momentum.

Secondly, an incoming single-slit state does not remain unchanged on passage through the wire system, but is instead detected on a set of locations expected of a joint eigenstate of projectors onto periodic sets in position and momentum space. The additional intensity peaks are found, such that each peak is separated by the same distance from its immediate neighbours as the two slits in the aperture. This is compatible with the interpretation that the single-slit state was projected onto an approximate joint eigenstate of spectral projections of position and momentum on periodic sets. Indeed, the original single-slit state, being already localised on a periodic set, has been changed into a state that is a good approximation to a joint eigenstate through the projective action of the wire grating. 

The fact that a single-slit input state is affected by the wire grating in such a way that the 
detected output state is found to be localised in many periodically spaced intervals is a demonstration 
of the mutual disturbance of measurements of incompatible observables. The projector $\chi_A( Q)$ onto 
a state localised in the single-slit region $A$ is not compatible with the projector $\chi_B( P)$ onto a state 
localised in the set $B$ of intervals in momentum space defined by the gaps in the wire grating or its idealised 
substitution by a periodic set. Consequently, a state originally prepared to be localised in $A$ is changed 
by the projective action of the wires so as to be less well localised in $A$ and instead localised in a 
periodic set. 

In this way the present experiment serves as a beautiful, new demonstration of complementarity that 
complements the existing illustrations. Usually one considers a perfect interference setup and then shows
how the interference pattern is degraded by the introduction of a path-marking interaction with a probe system 
storing (partial) path information. Here one starts with a perfect path-marking setup which then, by 
introducing the wires, is changed into an interference experiment, degrading the accuracy of path 
determination.

Finally, we used a construction of a specific class of joint eigenstates of periodic sets of position and momentum, which showed that in an idealised experiment with periodically placed slits and wires one can input such states that would propagate entirely unchanged through the setup, so that the presence of the interference pattern would be established without disturbing the quantum state at all. The work of Corcoran and Pasch \cite{Corcoran2004} suggests that the construction of realistic approximations to such quantum states is possible experimentally as well.

It is interesting to note that the work of \cite{APP} has been developed further in \cite{AR} and \cite{Tollaksen2010}.
Modular (periodic) momentum variables are introduced there since they are found to be sensitive to relative phase shift
of spatially non-overlapping partial wavefunctions and are thus indicators of the disappearance of interference fringes due to a path measurement. By comparison, here we are concerned with the diminishing path knowledge from creating an interference pattern, for which we found periodic characteristic functions of $Q$ and $P$ particularly useful. The cited work also introduces uncertainty relations allowing a quantitative description of the trade-off between path knowledge and quality of interference. 
It seems that there is an intimate connection between these uncertainty relations involving modular variables and 
trade-off relations between the overall width of a wavefunction and the fine structure of its Fourier transform that were 
formulated in \cite{UH}; an application of the latter uncertainty relation to the present experiment and a comparison
with the uncertainty relation for modular variables are work in progress.

To summarise, we have shown that it is appropriate to view the experiment reported in \cite{Afshar2007} as a preparation 
procedure for approximate joint eigenstates on periodic sets of position and momentum, whatever 
the input state. The validity of this interpretation {can be} supported by numerical simulations of the experiment 
and variants of it (with different numbers and thickness of the wires) \cite{Biniok2013}.

\section*{Acknowledgements}
We wish to thank Stefan Weigert for pointing out Ref. \cite{Corcoran2004} to us and for valuable comments and suggestions
on a draft version of this paper. Thanks are also due to Tom Bullock and Leon Loveridge for a critical reading of the manuscript. We are grateful to P\'{e}rola Milman and an anonymous referee for pointing out Ref. \cite{APP}.

\section*{Appendix A: Determination of the momentum distribution via late-time position measurement}
From classical optics it is known that the interference pattern of a wave passing through a double slit can be described by the Fourier transformed aperture profile. 
Additional analysis is necessary to justify the same application in quantum mechanics. 
In particular, it is required to show that after free evolution the position representation of the state $\psi_t$ 
at location (ii) is, 
up to scaling, approximated by the momentum representation of $\psi_0$, at the aperture at (i):
\begin{equation*}
\psi_t  \propto \widetilde{\psi}_0\quad (\text{approximately}).
\end{equation*}
We give a simple `rough and ready' argument here to show how this approximation can be obtained.
The solution of the Schr\"{o}dinger equation for free time evolution is given by
\begin{equation*}
\psi_t (x)=\sqrt{\frac{m}{2\pi i  t}} \int_{-\infty}^{+\infty}\psi_0 (x') \; \mathrm{exp}\left(i\frac{ m (x-x')^2}{2 t}\right) dx'.
\end{equation*}
With the limits of integration bounded by the apertures, the actual integration takes place from $-(T+a)/2$ to 
$(T+a)/2$. In the 
limit of large $t$ then, the term depending on $(x')^2$ in the exponential can be neglected to a good 
approximation, because 
it is bounded by the finite dimensions of the aperture.
\begin{equation*}
\psi_t (x) \approx \sqrt{\frac{m}{2\pi i  t}} \int \psi_0 (x') \; \mathrm{exp}\left(i\frac{m x^2}{2 t}\right) \mathrm{exp}\left(i\frac{ m x}{ t}x'\right) dx'
\end{equation*}
After trivial rearranging, the desired expression is obtained.
\begin{align*}
 \psi_t (x)  & \approx \sqrt{\frac{m}{i  t}}\; 
 e^{i\frac{m x^2}{2 t}}\frac{1}{\sqrt{2\pi}} \int \psi_0 (x')\;e^{i \frac{m x}{t}x'}
dx' \\
 & \approx \sqrt{\frac{m}{i  t}} \; \mathrm{exp}\left(i\frac{m x^2}{2 t}\right)\widetilde{\psi}_{0}\left(\frac{m}{t}x\right)
\end{align*}
The parameter $t$ can be eliminated using $\frac{p_z}{m}t= L$, with the distance to the lens $L=0.55 m$, 
where $p_z$ denotes the
longitudinal momentum component. In doing so, the limit 
of large $t$  becomes a limit of large distance $L$ in relation to the aperture size. Considering the dimensions 
of the setup used in \cite{Afshar2007}, where the centre-to-centre separation of the two pinholes is $0.25 mm$, 
this is reasonable. Furthermore, as $p_x/p_z$
will be small given these dimensions, we can also substitute $p_z$ approximately with the magnitude of the 
mean momentum, $p_0$
so that for  the mean wavelength $\lambda$ of the packet we can use the value $\lambda=2\pi/p_0\approx 2\pi/p_z$, and so 
$t\approx  mL\lambda/(2\pi)$. This gives the intensity as
\begin{equation*}
|\psi_t (x)|^2  \approx  {\frac{2\pi}{L\lambda}}\; \left|\widetilde{\psi}_{0}\left(\frac{2\pi}{L\lambda}x\right)\right|^2
\end{equation*}
Hence, measuring the interference pattern at location (ii) by determining the distribution of position $Q$ 
constitutes a measurement of
a scaled momentum observable with respect to the input state $\psi_0$. 
We can express this in terms of 
the spectral measures of $Q$ and $P$:
\[
\big\langle\psi_t \big |  \chi_{_{L\lambda Z/(2\pi)}}(Q)\psi_t \big\rangle\approx\left\langle\psi_0|\chi_{_Z}(P)\psi_0\right\rangle,
\]
where $Z$ is any (Borel) subset of $\mathbb{R}$. The separation of the wires in Fig.~\ref{fig:Diagram} 
was indicated as
being proportional to $2\pi/T$; the above consideration gives the separation in spatial dimensions as $L\lambda/T$.

\section*{Appendix B: Square-integrability}

{The relation (\ref{diff3}) may be used to define a wavefunction $\psi$ via equations \eqref{psi}, \eqref{psitilde}
for square-integrable $W,M$ if $W$ vanishes outside an interval of length strictly less than $T$,
because then the square-integrability condition is met, i.e. if the $L^2$-norm of $||\psi||_{2}$ is finite:}
\begin{widetext}\begin{align*}
||\psi||_2 & =  \int_{-\infty}^{+\infty} \left| W \ast (\Delta_T \cdot M)(x) \right|^{2} dx  \\
 & =  \int  \overline{W \ast \left(\sum_{n=-\infty}^{\infty} \delta((\cdot)-nT) \cdot M\right)(x)} \; 
 W\ast\left( \sum_{n'=-\infty}^{\infty} \delta((\cdot)-n'T) \cdot M\right)(x) \; dx \\
 & =  \int  \overline{W \ast \left(\sum_{n} \delta((\cdot)-nT)M(nT)\right)(x)} \; 
 W \ast \left( \sum_{n'} \delta((\cdot)-n'T)M(n'T) \right) (x)\; dx  \\
 & =  \int  \sum_{n}{\overline{W(x-nT)}\,\overline{M(nT)}} \; \sum_{n'}{W(x-n'T)M(n'T)} \; dx  \\
 & =  \sum_{n}{\left|M(nT) \right|^{2} \int \left|W(x-nT)\right|^2\, dx}
 =\left\|W\right\|_2^2 \sum_{n}{\left|M(nT) \right|^{2}}
\end{align*}\end{widetext}
{The last line is obtained due to the localisation property of the function $W$, which entails that $\overline{W(x-nT)}W(x-n'T)=0$
if $n\ne n'$.}
The square integrability of the Fourier transform $\widetilde\psi$ is ensured by the Fourier-Plancherel
theorem.

\section*{Appendix C: No delta functions beyond this point}

The construction of Section \ref{Sec:eigenfn} involved delta functions; here a different approach 
is presented that is mathematically rigorous without going into the theory of distributions. 
While similar to \cite{Reiter1989}, the result here is more general. 

Starting by choosing a square-integrable function $W$ with support strictly within the interval $(-T/2,T/2)$, 
we define a periodically-supported function $\psi$ as
\begin{equation} \nonumber
\psi(x) = \sum_{n=-\infty}^{\infty}{c_{n}\,W(x-nT)}.
\end{equation}
For each $x$, the sum contains exactly one term, hence the series converges pointwise. The coefficients $c_n$ are to be determined by further constraints below; here we note that given the square integrability 
of $W$, $\psi$ is square integrable if and only if the $c_n$ are square-summable. This entails that the series also converges in norm.
Note that
\begin{equation} \nonumber
\mathrm{supp}\; \psi = \bigcup_{n=-\infty}^\infty \mathrm{supp}\left(W +nT\right).
\end{equation}
Computing the Fourier transform yields
\begin{align} \nonumber
\widetilde{\psi}(k) & = \int_{-\infty}^{\infty}{\sum_{n=-\infty}^\infty{c_{n}\,W(x-nT)}\;e^{i \, k \, x}\;dx}
\\ & = \sum_{n=-\infty}^\infty c_{n}\;e^{i \, k \, nT}\;\widetilde{W}(k).
\end{align}
The coefficients $c_n$ represent the coefficients of a Fourier series expansion of a periodic function 
$\widetilde{M}_p$ with period $2\pi/T$:
\begin{equation} \nonumber
\widetilde{M}_p(k) = \sum_{n=-\infty}^\infty c_{n}\;e^{i \, k \, nT}.
\end{equation}
Let $\widetilde{M}$ be a function that is  supported  inside the interval $[-d,d]$ where $0<d<\pi/T$. We can then specify
$\widetilde{M}_p$ -- and hence the coefficients $c_n$ -- so that
\[
\widetilde{M}_p(k)=\sum_{n=-\infty}^\infty\widetilde{M}\left(k-\frac{2\pi}Tn\right).
\]
This function is supported in a periodic set,
\[
\mathrm{supp}\;\widetilde{M}_p\subseteq \bigcup_{n=-\infty}^\infty \left[ \frac{2\pi}Tn-d, \frac{2\pi}Tn+d \right].
\]
We thus have that
\[
\widetilde{\psi}(k)=\widetilde{M}_p(k)\;\widetilde{W}(k).
\]
A simple calculation shows that $\widetilde{M}$ is square integrable if and only if the $c_n$ are square summable.
As noted above, this condition is equivalent to $\psi$ being square integrable. With such a choice of $\widetilde{M}$
we can also see directly from the last formula that $\widetilde{\psi}$ is square integrable, in line with the
Fourier-Plancherel theorem.

\end{document}